\documentclass{article} % For LaTeX2e
\usepackage{iclr2026_conference,times}

\usepackage{amsmath,amsfonts,bm}

\def\eqref#1{equation~\ref{#1}}
\def\1{\bm{1}}

\DeclareMathAlphabet{\mathsfit}{\encodingdefault}{\sfdefault}{m}{sl}
\SetMathAlphabet{\mathsfit}{bold}{\encodingdefault}{\sfdefault}{bx}{n}

\usepackage{hyperref}
\usepackage{url}
\usepackage{xurl}
\usepackage{tikz}
\usetikzlibrary{calc}
\usepackage{graphicx}
\usepackage{rotating}
\usepackage{booktabs}
\usepackage{array} 
\usepackage{makecell}

\title{Formalising Human-in-the-Loop:\\Computational Reductions, Failure Modes, and Legal--Moral Responsibility}

\author{Maurice Chiodo \\
Centre for the Study of Existential Risk \\
University of Cambridge \\
Cambridge, United Kingdom \\
\texttt{mcc56@cam.ac.uk}
\And
Dennis Müller \\
Institute of Mathematics Education \\
University of Cologne \\
Cologne, Germany\\
\texttt{dennis.mueller@uni-koeln.de}
\And
Paul Siewert \\
Department of Computer Science and Technology \\
University of Cambridge \\
Cambridge, United Kingdom \\
\texttt{pks50@cam.ac.uk}
\And
Jean-Luc Wetherall \\
DeepFin Research \\
St.~Helier \\
Jersey, Channel Islands\\
\texttt{jean-luc@deepfinresearch.com}
\And
Zoya Yasmine \\
Faculty of Law \\
University of Oxford \\
Oxford, United Kingdom\\
\texttt{zoya.yasmine@law.ox.ac.uk}
\And
John Burden \\
Leverhulme Centre for the Future of Intelligence \\
University of Cambridge \\
Cambridge, United Kingdom \\
\texttt{jjb205@cam.ac.uk}
}

\newcommand{\turing}{involved interaction}
\newcommand{\many}{endpoint action}
\newcommand{\trivial}{trivial monitoring}
\newcommand{\Turing}{Involved interaction}
\newcommand{\Many}{Endpoint action}
\newcommand{\Trivial}{Trivial monitoring}

\iclrfinalcopy

\begin{document}

%PLEASE PLACE KEYWORDS HERE (3-6):
%Human-in-the-loop, Automated decision making system, Human oversight in sociotechnical systems, Oracle machine, AI safety, Trustworthy AI

\maketitle

\begin{abstract}
We  use the notion of oracle machines and reductions from computability theory to formalise different Human-in-the-loop (HITL) setups for AI systems, distinguishing between trivial human monitoring (i.e., total functions), single endpoint human action (i.e., many-one reductions), and highly involved human--AI interaction (i.e., Turing reductions). We then proceed to show that the legal status and safety of different setups vary greatly. We present a taxonomy to categorise HITL failure modes, highlighting the practical limitations of  HITL setups. We then identify omissions in UK and EU legal frameworks, which focus on HITL setups that may not always achieve the desired ethical, legal, and sociotechnical outcomes. We suggest areas where the law should recognise the effectiveness of different HITL setups and assign responsibility in these contexts, avoiding human `scapegoating'. Our work shows an unavoidable trade-off between attribution of legal responsibility, and  technical explainability.  Overall, we show how HITL setups involve many technical design decisions, and can be prone to failures out of the humans' control. Our formalisation and taxonomy opens up a new analytic perspective on the challenges in creating HITL setups, helping inform AI developers and lawmakers on designing HITL setups to better achieve their desired outcomes.
\end{abstract}

\section{Introduction}\label{intro}

Human-in-the-loop (HITL)---the practice of embedding human oversight into a computational machine, such as an Automated Decision Making System (ADMS) or AI system---is frequently invoked as a safeguard to ensure AI safety and accountability~\citep{Green.2022}. We show that the effectiveness of HITL for satisfying both safety and regulatory requirements hinges critically on the specifics of the ADMS's design and implementation. We present a novel formalisation of HITL setups as computational reductions (\S \ref{HITL reductions}). We analyse those  corresponding to total functions, many-one reductions, and Turing reductions, showing that each leads to vastly different safety outcomes. Existing legal frameworks, such as Article 22 of~\citet{GDPR.27.04.2016} and Article 14 of~\citet{EUAIAct.12.07.2024} only consider HITL under a simplistic view, and do not account for the variety of setups we show possible. And so, we clarify the various outcomes (\S \ref{reductions}), failure modes (\S \ref{Failure modes}), and legal and moral responsibilities (\S \ref{legal-moral}) associated with each of our setups. We further demonstrate an inherent, ever present tension when using HITL, whereby setups with increased human interpretability and explainability hinder clear attribution of responsibility, and vice versa (\S \ref{trade-off}). Ultimately, HITL is no panacea to the problem of ADMS safety, and we identify six concrete suggestions for its effective use (\S \ref{conclusion}).

Although introducing HITL is increasingly a strategic choice to improve sociotechnical systems~\citep{Grnsund.2020}, our focus is not on optimising the average-case performance, but rather on ensuring safety, reliability, and robustness. We are primarily concerned with preventing ADMS-related harms by recognising that different degrees of human involvement in HITL setups carry varied technical and ethical implications that should be recognised by the law. Our perspective therefore asks when a HITL setup represents genuine and meaningful human control of an ADMS, and encompasses its moral, legal and political ideals of keeping `social control' over technology~\citep{Abbink.2024}. Our manuscript responds to the question of when, if ever, should we use HITL? And how can we select HITL setups in order to minimise harm and risk and thus ensure safety (as defined in \ref{ap_harms})? (cf.~Pillar 1 in \citet{chiodo2023manifestoresponsibledevelopmentmathematical}).

Much terminology exists for human--ADMS interactions. HITL setups have humans actively integrated into the ADMS's deployment cycle (we do not cover human input in training or development). Human-on-the-loop (HOTL) have humans acting as supervisors, intervening only when necessary. Setups where the ADMS operates without direct human  intervention are termed Human-out-of-the-loop (HOOTL). Human-in-Command (HIC) has humans determine the high-level functioning of the ADMS. Meaningful Human Control (MHC) was introduced to study what influence a human should have on the execution of an action and what levels of cognitive and moral awareness they need. These terms, and relevant literature, are detailed in \ref{ap_existing terms}. Our work demonstrates how different HITL setups can be generalised and formalised using  computability theory, unifying these related but disparate concepts. Such formalism creates a  general framework to analyse HITL setups across varied contexts, and provides a consistent way for regulators to examine HITL involvement where required by law, thereby also reducing the ability for companies to introduce tokenistic HITL setups.

Though intended to protect society, HITL setups can end up protecting the ADMS instead, with the human  becoming  the `moral crumple zone'  taking on accountability and enabling potentially faulty ADMSs to persist \citep{Elish.2019}. Hence, a closer scrutiny of HITL setups and their failure modes is necessary. The lack of a well-defined formal meaning of what HITL can and does involve, and why HITL may be beneficial in sociotechnical systems, is a recognised problem. And yet, having a HITL setup is frequently presented as a critical safety measure, even in high-risk domains (detailed in \ref{ap_existing concepts}). Recognising the different degrees to which humans are, and should be, involved in HITL setups, we present a classification of these using computational reductions. Our primary contributions are threefold. In \S \ref{reductions} we give a novel formalisation of HITL setups which unifies various known HITL categorisations and identify a fundamental distinction within HITL setups. In \S \ref{Failure modes} we give a taxonomy of failure modes and ethical concerns for HITL setups and find they are correlated with our distinction. And in \S \ref{legal-moral} we consider what HITL setups the law requires and how responsibility is assigned when they fail. In doing so, we contextualise the failure modes and categories of HITL setups to highlight ways laws can be improved to ensure safety and efficacy. Our work also uncovers an unavoidable trade-off between responsibility and explainability in HITL setups.

\section{Computational Reductions for HITL Setups}\label{reductions}

This section aims to characterise  computationally distinct HITL setups, using the concept of reductions. An \textbf{oracle machine} is a deterministic automaton $T^\bullet$ on a fixed language $W$ with a work tape, an extra `oracle tape', and some states marked as `oracle states'. For any function $f: W \to W$, $T^\bullet$ gives rise to a `machine with oracle' $T^f$, whose computations proceed as usual, but which whenever it enters an oracle state has the content $w \in W$ of the oracle tape (instantly) replaced by $f(w)$; whenever $T^\bullet$ halts, it outputs the oracle tape. This definition is reminiscent of the notions of $\exists$-non-determinism and random automata, in that $T^\bullet$ by itself defines, for any input, a computation tree that contains all possible behaviours depending on the behaviour in the oracle states (respectively, the non-deterministic states or the random states). The difference is in semantics, as the result of one computation is not defined by some general property of the computation tree, but by `plugging in' $f$ and thus resolving all the choices with $f$. A discussion of oracle machines can be found in~\citet[Chapter III, Section 1]{soare87}; we adopt here a variation of the terminology from~\citet[Section 2.4]{melkebeek}. Our analysis assumes the oracle $f$ is fixed;  even then, the computational power of $T^f$ may vary drastically as we vary $T^\bullet$. The most salient distinction for us is that between machines which only call $f$ once, and those which call $f$ an unbounded number of times. These correspond to two different notions of `reduction' between functions $f$ and $g$: if there is an oracle machine $T^\bullet$ such that $T^f$ computes $g$, one says $g$ is \textbf{Turing-reducible} to $f$. If moreover $T^\bullet$ is such that it halts immediately after its first call of the oracle, then one says $g$ is \textbf{many-one reducible} to $f$. 

Observe that an oracle call may be rather trivial. $T^\bullet$ might  ignore the content of the oracle tape, in which case the choice of $f$ does not matter. Even if $T^\bullet$ reads/writes from the oracle tape, its computational result might not depend on these steps. However, our concern lies with the \textit{function} defined by $T^{f}$, not the machine itself (even though one could examine it to increase computational transparency). Therefore, we are interested more so in what we define as a \textbf{real query}, which we say occurs at an oracle call if a fork exists at that point in the computation tree \textit{and} not all branches have the same set of possible outputs. The former of these ensures the oracle can still have some meaningful impact on the computation, and the latter ensures we avoid an `all roads lead to Rome' scenario where the machine can take multiple computational paths all leading to the same output.

\subsection{HITL setups as a computational reduction}\label{HITL reductions}

We propose describing human--machine systems in terms of this formalism. Any human--machine computational process uses algorithmic components at some points and human interventions at others. We hence conceive of a HITL setup as a type of Human-Based Computation~\citep{HBComp}, in  that the machine assigns particular tasks or problems to one or more humans to solve. Our perspective is different to how the term is typically used. We are not interested in large-scale human knowledge, but rather in the specialised knowledge of individuals. And the human is not the subordinate labourer of the machine or a simple assistant, but rather is in symbiosis with the entire system. We now examine this to see what computational synergy can exist between human and machine.

The architecture of a decision-making process including the algorithmic specifications corresponds to the oracle machine $T^\bullet$. The function $f$ is provided by the human: at certain points in the process, human \textit{judgment} on values and data enters the computation, which corresponds to $T^\bullet$ calling the oracle, and which we denote as a \textbf{human query}. We do not assume the human is `correct' (see \S \ref{Failure taxonomy} for HITL failure modes), or even deterministic (for details on how we see the human's role, see \ref{ap_human-nondet}). Here we include human interventions that are otherwise uninvited by the machine; computationally, it makes no difference what precipitated the human input (though in reality this still needs to be considered for evaluating safety). In response to a human query, we also allow the human to write an `emergency stop' symbol to the oracle tape, denoted by \textbf{!}; whenever the machine reads ! on the tape it halts immediately with no output. Note that the possibility to halt after a query with no output does not count towards the set of computational outcomes when determining if a query is real.

If an oracle machine is set up to never ask a human query, and humans have no way to intervene in its  operation, this corresponds to a HOOTL  from \S \ref{intro};  we will not cover this further. If an oracle machine is set up to ask human queries, but none are real queries, then we denote this as a \textbf{human trivially monitoring the loop}, abbreviated to \textbf{\trivial}. This corresponds to a HOTL setup from \S \ref{intro}, but where the human is only able to stop the process. Computationally, in a \trivial{} setup, the  human is not affecting the computational steps of the machine in any meaningful way, and serves only to prevent the machine from continuing its computation (though this is still a crucial safety role, as argued by~\citet[p.~448]{Crootof.2023} when discussing human sign-off requirements by `positioning human(s) as end-of-the-loop gatekeepers'). This means that $T^\bullet$ defines a total function independent of the human. The human can only decide whether to terminate the machine before it completes its computation (an `ignorance is bliss' scenario). If an oracle machine is instead set up to ask precisely one real human query and then immediately halt, giving the answer to that real query as its output, then we denote this as a \textbf{human at the end of the loop}, abbreviated to \textbf{\many}. Here, the machine does some computation itself, then hands over to the human to perform the rest. If the oracle machine is set up to ask more than one, and potentially unbounded, real queries (i.e., a number not a priori bounded before the machine starts computing), we denote this as a \textbf{human involved in the loop}, abbreviated to \textbf{\turing}. Practically, this is most resembling the scenario of human and machine working together, where the machine and human engage in a game of computational ping-pong: the machine does some work, hands over to the human who does some more and then hands back to the machine, and so forth. For example, a `creative' collaboration with a Large Language Model (LLM) where the human prompts the LLM to produce an initial response (or ask a question) which the human uses to determine the next detail they give the LLM, and so on. We illustrate all these  setups in \ref{ap_ping-pong}.

Thus, in an \textit{\turing{}}, the oracle machine Turing-reduces the computation to the human, but \textit{does not} many-one reduce it. In an \textit{\many{}}, the oracle machine many-one reduces the computation to the human, but it \textit{does not} define a total function. And in \textit{\trivial{}},  the human can terminate the machine, but not otherwise influence its computation. As  explained in \ref{ap_intermediate}, we choose not to study the myriad of intermediate reduction types, as \many{} and  \turing{}  give the `simplest' and `most complicated' ways a human can interact with a machine. 

As an example, consider a route-planning machine in a HITL setup with the human driver of a car. It may demonstrate \trivial{} (presenting one route for the human to accept or reject), or \many{} (presenting the human several routes to choose their preferred one), or \turing{} (where the human is fully involved in the  process, from early choices of where/when to travel, up to the latter optimisation of different route types, number of stops, etc.); \ref{ap_routes} expands on this. With growing reliance on human queries, the agency of the human and the access of the machine to human values increases, resulting in decisions becoming more adjusted to human needs.

\subsection{Why are we considering these reductions?}\label{computable trade-off}

Computationally, a Turing reduction between functions is viewed as weaker than a many-one reduction, because we fix the computational problem first, and then ask what oracles that problem reduces to. Our considerations with HITL setups (elaborated in \ref{ap_involvedbest}) are the opposite; we take a \textit{fixed oracle} (a human), and see which problems can be solved with that oracle (in \ref{ap_human-nondet} we explore the sense in which the human is `fixed'). Thus, in terms of constructing optimal HITL setups, an \turing{} allows us to solve the most problems with a given human. We thus propose that the HITL setups with the greatest potential to achieve agency, alignment, safety, transparency, and thus overall reliability, are those with the greatest reliance on human input, i.e.,  \turing{}s (detailed further in \ref{ap_involvedbenefits}). The more real queries made, the more we have a human \textit{in} the loop (rather than a HOTL). In existing work,  \citet{Adersen2024Design} describe 10 \textit{design patterns} for HITL setups, with three centred around HITL during deployment/inference. The first is the `Recommendation System' where a human  makes the final decision based on the machine output. The second is `Active Moderation' where, for a set of tasks, humans perform them in the cases where the machine cannot do so reliably. The third option `Thumbs up or Thumbs down' sees the human either accept or correct the machine output. These are all \many{}s or \trivial{}s. Notably, none of~\citet{Adersen2024Design}'s patterns describe interactions where the task is aborted rather than corrected---effectively requiring the human to be as capable as the machine at the initial task (rather than being able to `merely' recognise an error). In \S \ref{legal-moral} we  connect this to Meaningful Human Control.

In a \trivial{} setup, the human is not involved in the computation between the machine starting and finishing its work. And so in an opaque `black box' computational process, such as many ADMSs, one may have no idea how it was carried out. However, one can begin to `unmask' the black-box if the machine asks real queries, as each precipitates a human-interpretable question giving \textit{some} information about what the machine is currently `doing' at that point in the computation. The more real queries, the more effective this unmasking is. In an \many{} setup, the machine reveals \textit{one} computational step at the very end, giving \textit{some} insight into its workings. And in an \turing{} setup, there may be many real queries, revealing insight at many points throughout the computation. Of course, there may still be `black-box' computation between these real queries. But rather than one \textit{big} black box, the process appears as many \textit{smaller} black boxes connected at the points where a human provides input. The machine--human `ping-pong' can be viewed as a \textit{chain of computations}, illustrated in \ref{ap_black box}. Thus, in an \turing{} setup we argue that these smaller black boxes increase \textit{explainability} of what went into determining the final output. This chain is relevant again in \S \ref{trade-off} where we discuss responsibility within HITL setups.

\subsection{Determining HITL setup types in practice}\label{practicalities}

So how do we \textit{identify} a HITL setup as \trivial{}, \many{}, or \turing{}, and why use such formalism? \ref{ap_HITLtypes} shows this is important, yet difficult. \textit{Technically}, showing non-existence of a `simpler' reduction is hard, as it cannot be done by example but instead requires deep analysis of the computational process. \textit{Legally}, showing the human will do something meaningful is hard, as asking the human a series of `pointless' questions may violate the legal principles of `meaningful'. And \textit{morally}, showing the HITL setup is not simply a facade masking a less-involved process is also hard, as one needs to ensure the machine does not simply ignore answers the human gives (or worse: inverts answers).  Overall, our classification of HITL setups into \trivial{}, \many{}, and \turing{} outlines what is \textit{possible} with HITL. But identifying such setups, from the technical, legal, and moral perspectives above, is another challenge altogether.

Our route-planning example in \S \ref{HITL reductions} raises  questions about the \textit{practicalities} of HITL setups.  In the \many{} scenario, the machine could have instead asked the human a fixed finite series of questions such as `Prioritise speed or  efficiency?', `Maximum acceptable distance between fuel stations?', etc., and done computation between each, to find a route. This would  \textit{appear} to be an \turing{} (actually a bounded truth table reduction; see \ref{ap_boundedtruth}). However, these could be encapsulated by instead asking the human to chose from a (long) list presented at the end; an \many{}. 
But in \textit{practical} terms, asking a human $20$ binary questions is much more feasible than presenting $2^{20}$ options.  Computational aspects are \textit{one} consideration in HITL setups; the abilities and limitations of humans are another---they do not operate like abstract oracles. But, to  understand how human input feeds into a HITL setup, and `design out' the failure modes presented in the next section, we must have first understood how the human is involved from a computational perspective.

\section{HITL Failure Modes}\label{Failure modes}

\subsection{Characteristics of effective HITL setups, and taxonomy of failure modes}\label{Failure taxonomy}

According to~\citet{10.1145/3630106.3659051}, necessary and sufficient conditions for effective (general) oversight of AI are given when the human overseer has 1) an adequate understanding of the system, 2) the self-control to act on their judgment, 3) the power to intervene effectively, and 4) intentions aligned with the oversight goals. They, and others, note that successful oversight depends on many user-specific attributes, including technical skills, domain knowledge, general attitudes towards technology, especially those related to interpreting AI outputs, as well as attributes of the human--machine setup (cf.~\citet{10.1145/3630106.3659051, Sudeeptha.etal.2024, langer2025test}). The arguments presented in the literature remain at a comparatively abstract level, and do not distinguish by HITL setup. Overall, as our reductions in \S \ref{HITL reductions} show, the functioning of HITL setups does not rely on the design of the machine alone, nor  on the characteristics of the human overseer, but on how they are put into relation with each other.  This complex sociotechnical relationship allows us to turn the theoretical and empirical insights from the literature into practice, and analyse potential failure modes.  We  present our own taxonomy centred around five main failure categories, giving a partial breakdown of each:

\textbf{1. Failure of the machine components.} These include: Unexpected inputs or outputs, problematic machine evolution or self-adaptation, biased or erroneous outputs, and other unexpected  behaviour.
\\\textbf{2. Failure of the process and workflow.} These include: Insufficient power of the human, insufficient self-control/independence,  insufficient reaction time,   unrealistic expectations,  delayed notification,   insufficient support, and other process and workflow failures.
\\\textbf{3. Failure at the human--machine interface.} These include: Incomprehensible or incomplete outputs, complex or poorly designed user interface, insufficient training, and other epistemic failures.
\\\textbf{4. Failure of the human component.} These include: Cognitive bias,  automation bias,  fatigue,  incongruous intentions, stress or overload, lacking courage, and other human-centric failures.
\\\textbf{5. Exogenous circumstances.} These include: Unreasonable laws, unreasonable societal expectations, inappropriate workplace requirements, and other external pressures.

The empirical rationale and literature behind these categories, along with an extended table of failure modes, is given in \ref{ap_HITLlit}. The categories are ordered by `amount of human-ness', from purely digital failure (no human-ness), to failure of social pressure and wider society (vast human-ness). Our taxonomy covers HITL setups in general computational machines, of which ADMSs are  an example.

\subsection{Connections to our HITL setup types}\label{connections}

Setups configured for \trivial{}, where the human cannot provide computationally meaningful input beyond `proceed' or `halt', may be particularly susceptible to failure modes related to the human component. The human's comparatively passive role means that automation bias, fatigue, or simply lacking the courage or perceived authority to halt a process become significant risks. This setup can also mask process and workflow design failures, such as the human having insufficient power to truly intervene or having unrealistic expectations about the level of oversight provided. Failures originating in the machine components themselves might also proceed unchecked, as the monitoring human may lack the mechanism or mandate for closer scrutiny. 

An \many{} setup relies on a single, critical human input after the machine has performed its computation, concentrating failure risks around that specific interaction point. Human--machine interface failures become critical: if the machine presents incomprehensible or incomplete outputs, or if the interface is poorly designed, the human may be unable to make informed decisions. Similarly, human component failures like cognitive bias can influence their single judgment with no backup checks present. Failures in process and workflow design, such as insufficient reaction time allowed for the human or delayed notification that input is required, are also highly relevant here. 

In an \turing{} with potentially many queries, failure modes quickly become more complex. The back-and-forth nature of the process makes larger design failures (e.g., unclear roles) and human--machine interface issues more pronounced. While increased interaction might help the human catch machine failures (and  vice-versa), and  break down a vast set of options into a manageable list of choices (see \S \ref{practicalities}),  it may also lead to unexpected machine behaviour if the machine adapts to the user in unintended ways. Human failures such as fatigue or stress from prolonged interaction, or the accumulation of cognitive biases across multiple decision points, can additionally degrade performance over time. Furthermore, the complexity involved can hide superficial engagement with the machine outputs when quantitative interaction is mistakenly understood as qualitative interaction: many smaller human--machine interactions may not sufficiently change the machine's output. Overall, while a HITL setup requiring \turing{} gives the human the most power to intervene, it is also the setup which is most complex, thus leading to more nuanced failure mechanisms.

In summary, different failure categories may be more likely for different HITL setups, and each category can be realised through different failure modes. We believe this taxonomy of five failure categories is able to capture \textit{most} HITL failures. However, the list below each category should not be seen as exhaustive, but rather reflects a selection from the wider literature, as covered in \ref{ap_HITLlit}. They are here to illustrate that one should not only consider HITL setups via reductions, but also by simultaneously grouping different failure modes into our categories from \S \ref{Failure taxonomy}. This provides a two-dimensional actionable picture of HITL setups focused on harm prevention. Akin to the oversight of general mathematical technologies \citep{chiodo2023manifestoresponsibledevelopmentmathematical, muller2025integrators}, ignoring entire categories is a likely way to failure, while ignoring individual list items can still lead to failures in specific contexts. In short, reductions and failure categories should always be considered together.

\subsection{Examples of HITL setup failures}\label{Uber example}

This taxonomy enables us to identify where and why real HITL deployments fail. We can apply it to several case studies of (failed) HITL setups, showing how these setups failed in relation to the above taxonomy. What comes to light from these is that failures of HITL setups are usually due to poor integration \citep{muller2025integrators}, and what is often put down to `technical failure' or `human error' can actually be avoided if proper integration is carried out (as happened with an ADMS security scanning setup at a sports stadium, where two firearms were brought through security and the problem was blamed on `human error'; see \ref{ap_MCG} for details).  To further illustrate the failure facets of \many{} HITL setups, \ref{ap_Notre} analyses the catastrophic fire at Notre-Dame Cathedral. Here, we turn to a fatal self-driving car collision involving a \trivial{} HITL setup:

As presented in~\citet{Hawkins.2018, Fitzsimmons.2018, NationalTransportationBoard.2019}, in March 2018 the first recorded incident of a pedestrian fatality from a collision with a self-driving car occurred. The car had a \trivial{} HITL setup created by Uber, with a human driver (an Uber employee) at the wheel poised to intervene and `take over' the driving at any point if they observed a problem with the self-driving mechanism~\citep[p.~8]{NationalTransportationBoard.2019}. In this HITL setup, the human operator monitored the ADMS responsible for autonomous driving and had the ability to intervene by braking and/or taking the wheel, but they did not have the ability to change the ADMS's decision-making in more complex ways. The human monitor was expected to perpetually `hover their hands above the steering wheel and foot above the brake pedal' (ibid., p.~12) (unrealistic expectations), without ever touching them as that would disable the self-driving mode (ibid., p.~11). The car had been in self-driving mode for 19 minutes before the crash, with no tasks required of the human driver (ibid., p.~19) (fatigue, automation bias). The self-driving ADMS first identified the pedestrian 5.6 seconds before collision, as they were jaywalking across the road. However, as it had not been programmed to classify a jaywalker (ibid., p.~16), it spent the next 4.1 seconds misclassifying the pedestrian as `another vehicle', `bicycle', and `other' (ibid., p.~15) (unexpected inputs). The ADMS first predicted a collision 1.5 seconds before impact and began evasive action; at 0.2 seconds before impact the ADMS established that impact was inevitable, instigated a controlled slowdown, and gave the human driver the first warning notification of upcoming impact (ibid., p.~16) (delayed notification, insufficient reaction time). The human driver had, from 6 seconds before impact, been gazing down at the control panel (ibid., p.~18), allegedly watching their cellphone (ibid., p.~24) (incongruous intention). With a warning lead-in time of 0.2 seconds---approximately human reaction time---the human driver had virtually no time to take control; they took the wheel 0.02 seconds before impact, and the car then hit the pedestrian. The entire HITL setup was affected by a poor safety culture (ibid., p.~38) (insufficient support, other external pressures). In this example, the failure cascade spans across our taxonomy: (1) the ADMS components failed by not recognising the pedestrian, (2) the workflow failed as the human allegedly was not embedded in a supporting safety-culture and had insufficient reaction time and unrealistic expectations, (3) the human--machine interface setup failed by allowing an environment whereby the human could watch a video, and (4) the human failed by being inattentive. However, one key further failure of this setup arose post-incident, when the law failed to hold the company accountable in any way or form (corresponding to (5) from our taxonomy), which brings us the final aspect of our manuscript: responsibility.

\section{Legal--Moral Responsibility}\label{legal-moral}

We now narrow our focus to ADMSs. If a HITL setup fails and causes harm, the law and morality investigate what went wrong and who is at fault. The Uber case (\S \ref{Uber example}) highlights the challenges when legal frameworks intersect with HITL failure modes. In this case, both criminal and civil liability fell on the Uber \textit{employee}; the human operating the car. This was despite allegations that Uber's technology was flawed (i.e., not recognising the jaywalking pedestrian) and the company's poor safety culture (for  case details, see~\citet{stamp2024reckless}). In the final section of this manuscript, we consider relevant (EU, UK) legal frameworks concerning design setups and liability and point to ways that they can be improved by incorporating our formalisation of HITL setups. We also identify a trade-off between responsibility and explainability in choosing setups. Though, in all HITL setups where responsibility gaps emerge, we argue that there should be a more nuanced approach to liability, to avoid the scapegoating of humans to shield technology companies like in the Uber case.

\subsection{HITL setups and the law}\label{HITL law}

The UK and EU General Data Protection Regulation (the \textit{GDPR}, which only applies to \textit{personal} data processing (Article 4(1) GDPR)) and the EU AI Act have imposed requirements for human oversight mechanisms to be implemented for automated decision making and \textit{high-risk AI} (see Article 6 and Annex III of the EU AI Act). Both legal frameworks take a `by design' approach, requiring developers to embed certain safety mechanisms into the technical setup of their ADMSs before they are deployed. This means that the law does not only threaten legal liability when things go wrong (discussed later in \S \ref{scapegoat}), but it also mandates measures to \textit{prevent} harm from occurring. Given the focus of the GDPR and EU AI Act in relation to HITL, lawmakers have acknowledged that HITL setups are a crucial tool to prevent harm in high-stakes scenarios like biometrics, law enforcement, and employment (\citealp[Annex III]{EUAIAct.12.07.2024} and~\citealp[Article 22(1)]{GDPR.27.04.2016}).

Article 22(1) of the GDPR governs `solely automated decisions', generally prohibiting \trivial{} setups (as detailed in \ref{ap_SCHUFA}) that could cause legal effects (or similar) on individuals, like credit applications or e-recruiting practices~\citep[Recital 71, GDPR]{Vollmer.2023}. To go beyond \trivial{} and avoid falling within the scope Article 22 (which requires consent and other safeguarding requirements), companies need to implement \textbf{`meaningful oversight'}; see \citet[p.~19]{EuropeanCommission.2018}: currently understood as an \many{}. Similarly, the EU AI Act requires high-risk AI to be `designed and developed in such a way, including with appropriate human--machine interface tools, so that they can be \textbf{effectively overseen}' by individuals while in use~\citep[Article 14(1)]{EUAIAct.12.07.2024}. The wording `effective' and `meaningful' in both the GDPR and EU AI Act suggest that \trivial{} alone is insufficient to meet legal obligations. A  \trivial{} setup with no influence on the decision or computational outcome cannot be doing anything `meaningful' or `effective'. Current laws focus on the role of humans at a very late stage in the HITL setup, in either \trivial{} or \many{} setups and as \citet[p.~4]{sarra2024artificial} notes, `substantial human intervention in previous stages appears to be irrelevant'. But the law (and related guidelines) do not indicate what alternative HITL setups should be implemented (see \ref{ap_SCHUFA}). 

\subsection{Moving towards \turing{}}\label{moving}

Our computational classification of HITL setups provides a framework to compare the significance of the human's involvement, and their ability to reduce harms within each setup. More frequent human interventions have computational and ethical implications (\S \ref{computable trade-off}, \ref{connections}). We argue that stronger reductions involving at least some (and potentially unbounded) querying of the human should be required to implement `meaningful' or `effective' oversight as stipulated by the law. By setting out technical setups that align with legal requirements, developers are incentivised to design their ADMSs with greater human involvement, which can improve system safety. Beyond the oversight requirements, the EU AI Act and GDPR also prescribe certain safeguarding duties on the human, whereby they are expected to prevent or minimise risks to `health', `safety', `legitimate interests', and `fundamental rights'. Yet, in \trivial{} or \many{}, humans are not effectively enabled to perform their safeguarding duties because they may face a completely `black box output' from a machine (\S \ref{computable trade-off}). From such an output it could be impossible for the human to assess whether any rights or interests have been infringed, due to any inherent opaqueness of the ADMS's output and of the factors that influenced the decision.

By contrast, in \turing{}s, the human is actively enabled to meet their safeguarding duties. In a HITL setup where the ADMS asks many questions, the human can better assess what should or should not be factored into the ADMS's output to make fair decisions that do not infringe rights or interests (\S \ref{computable trade-off}). Thus, the human is empowered to scrutinise the ADMS's process. Further, \many{} or \turing{} setups enable the human to input their own information. Depending on what the ADMS asks, the human may be able to better align the output of the ADMS with human values, for example by confirming or denying the relevance of certain input factors, like religion, race, or sex. \ref{ap_safeguarding duties} gives an example of the benefits of stronger reductions (like \turing{}) with a recent legal case involving SCHUFA---whereby an automated credit scoring system with a (weak) \many{} HITL setup was deemed by the courts as acting as a \trivial{} setup (\ref{ap_SCHUFA}). As explained in \ref{ap_safeguarding duties}, by implementing an \turing{} (\S \ref{HITL reductions}) to break the automation bias of the human (\S \ref{Failure taxonomy}), SCHUFA could have prevented the slip back to \trivial{} and thus avoided violating Article 22 (\S \ref{HITL law}). Of course, \turing{}s are not a perfect solution as they still suffer all the potential HITL failure modes (\S \ref{Failure taxonomy}, \ref{connections}). But, in general, they put companies and humans in a better position to meet their legal and moral obligations to provide oversight as well as safeguard the decision subject's rights. But ultimately, without proper consideration for what actions the human can take \textit{in theory} (\S \ref{reductions}), and the ways in which human--machine interactions fail \textit{in practice} (\S \ref{Failure modes}), the human may have to be \textit{superhuman} to meet these safeguarding duties.

\subsection{HITL responsibility and explainability trade-off}\label{trade-off}

While an \turing{} may enable the human to positively impact and scrutinise the output of an ADMS, this comes with a trade-off which complicates legal and moral responsibility. `Responsibility gaps' refer to situations where the `black box' features of autonomous technologies, combined with the complexity of the sociotechnical system, obfuscate an immediate source of responsibility for the impact of an ADMS (assisted) decision~\citep{matthias2004responsibility}. Introducing significant human influence into a system has been posited as a way to reduce responsibility gaps when compared with systems with limited or no human influence \citep[p.~194]{sienknecht2024proxy}. Indeed, using our reductions framework, in a HITL setup with a weak reduction such as an \many{}, it may well be easier to directly link the impact of the ADMS with the actions of the human because they may have approved the ADMS's output, thus closing responsibility gaps. Conversely, in an ADMS with no human, `the ADMS' is responsible (though assigning responsibility to an ADMS opens up a web of distributed responsibility, where a number of parties may share moral responsibility). But, as we now show, the link between responsibility gaps and HITL is far more complex.

While we advocate for stronger setups (\turing{}s) to meet GDPR and EU AI Act obligations, within these it is immensely complex to determine how the human input(s) impact the ADMS output due to the human--machine entanglement (see \ref{ap_Explainability}). In \S \ref{computable trade-off} we described a chain of computation. But the longer this chain, the more obfuscated responsibility is, as it becomes harder to pinpoint which human and/or machine decision(s) had the most impact. While the human does have more  impact on the ADMS, the impact of these inputs, even if recorded (\ref{ap_Explainability}), on the ADMS's output remains potentially unknown. By contrast, it could be relatively simple to identify the human's impact within a setup like \trivial{} or \many{}, because they effectively make a  decision at the start  whether to use an ADMS, and a decision at the end whether to  use  its output and if so, how. Here, the impact of the human on the overall ADMS and its output is clear, unlike in an \turing{}. As such, we witness a trade-off within \turing{}s. On the one hand, they enable the recording of questions asked by the ADMS and the responses inputted by the human, increasing transparency and explainability of contributing factors (cf.~\S \ref{computable trade-off}). On the other hand, the human--machine entanglement erodes attribution of the human's impact on the ADMS's outputs, creating responsibility gaps from the complexity of identifying which decision point(s) led to system failure.  This is an unavoidable explainability--responsibility trade-off: a more explainable HITL setup with clearer intermediate computational steps obfuscates responsibility; a setup with clearer attribution of responsibility is far more `black-box'. This trade-off needs consideration when using the law  to motivate  building  better HITL setups, as the two can operate at cross purposes.

\subsection{Avoiding the HITL `scapegoat'}\label{scapegoat}

So far, we have identified omissions in the law's approach to moral responsibility gaps and the understanding of the technicalities of HITL setups. To avoid the human being treated as a scapegoat (like in the Uber case), we argue that regulations should provide guidance on HITL setups in terms of reduction types, to incentivise companies to design such setups more effectively. The GDPR and EU AI Act go some way to resolve these emerging responsibility gaps by imposing liability on the data controller or technology provider to ensure that ADMSs are \textit{designed in ways} that enable `meaningful' or `effective' oversight to allow the human to safeguard the rights of decision subjects. This manuscript contributes by showing \textit{how} this might be done from computational (\S \ref{reductions}) and practical (\S \ref{Failure modes}) perspectives. This is important because the onus is on the company, whether or not the human causes harm, to design setups effectively. Nevertheless, the scope of these legal frameworks is limited and claims may also arise in negligence law when harm has occurred. For example, the UK courts have previously departed from established principles in challenging cases involving `responsibility gaps' to compensate individuals who have been harmed but the cause of injury cannot be discerned (for  details see \ref{ap_Mesothelioma}). These cases involved workers developing deadly mesothelioma from exposure to asbestos fibres across multiple employers~\citep{Barker.2006}. There, liability was calculated by an amount relative to the proportion of exposure the claimant had at a given employer, even though they might not have actually caused harm directly. The principles underlying the court's treatment of the mesothelioma cases may provide inspiration for how we should address similar responsibility gaps in HITL setups with \turing{} (\ref{ap_Mesothelioma}). 

In the US, one already sees a tendency to hold the human liable for the failure of an ADMS~\citep{stamp2024reckless}. The authors are apprehensive about such liability approaches in the event of failure across all computational setups from \S \ref{HITL reductions}.\footnote{Unless the human has acted in ways that can be proved as deliberately malicious or seriously negligent. But even then, ADMS designers are responsible for implementing controls that prevent human malevolence and create fail safes and checks on how the human interacts with the ADMS.} It might seem like the most intuitive response, especially in \trivial{} and \many{} where the human often has  final control over the actions. Yet, as our taxonomy in \S\ref{Failure taxonomy} shows, failures arise for multifaceted reasons which the human might have limited control over. Even in \turing{}s, where moral philosophy has shown us that responsibility is more complex, the human should not be used as a complete `scapegoat' to shield companies from accountability for their contributions to failures. The mesothelioma-style cases provide a foundation for how the courts should respond to this which aligns with the computational realities of \turing{}s and the responsibility gaps that emerge at these intersections. Where the human also has onerous obligations to safeguard individuals and ADMSs are error prone, biased, and frequently act in unexpected ways, we need to avoid the human taking total liability for all failures.

\section{Conclusion and Suggestions}\label{conclusion}

Our analysis brings new clarity to the design of HITL setups by characterising them through computational reductions and complementing this formalism with an empirically motivated taxonomy of HITL failure modes. HITL setups involve complex sociotechnical decisions and are susceptible to failures beyond human control, which necessitates this joint perspective for designing effective and responsible setups. Our analysis connected these setups and failure mechanisms to existing limitations and omissions in the law. This allowed us to make suggestions for more refined rules surrounding HITL requirements, as well as identify a trade-off between responsibility attribution and technical explainability, and recommend that courts cautiously approach liability in these cases. We thus make the following suggestions for those  designing or regulating HITL setups:

1. Define the computational HITL setup, if possible aiming for more than \trivial{}.
\\2.  Avoid `bolting-on' HITL to existing workflows; they must be fully integrated into the process.
\\3.  Establish guidelines for meaningful human oversight that consider different HITL setups.
\\4. Ensure that expectations placed on humans in HITL setups match their competency. 
\\5. Implement  measures to prevent humans becoming `moral crumple zones' protecting  machines.
\\6. Understand the trade-offs between legal clarity and technical explainability, to inform more nuanced approaches to assigning liability in HITL failure cases.

If done poorly, a HITL setup can create a dangerous two-way deferral of responsibility between machines and humans. Humans may overly defer to machine computation, and machine designers may overly rely on humans for safety, all of which can lead to disastrous consequences. Integrating HITL is not a binary process; many ways exist, with different consequences. As a bad HITL setup may be just as harmful as no HITL setup, achieving a \textit{good} setup needs to be the objective.

\section*{Author Contributions}

The first author (Chiodo) led the project. Authors 2--5 (Müller, Siewert, Wetherall, Yasmine) are listed alphabetically by surname, and all contributed equally. The final author (Burden) oversaw the project. All authors contributed to drafting, writing, and editing.

\section*{Reproducibility Statement}

The vast majority of results and analysis in this work are of a completely theoretical nature, and thus can be  verified through further theoretical research. As outlined in  \ref{ap_HITLlit}, the initial empirical foundation of the taxonomy of \S \ref{Failure taxonomy} is derived from a series of ethics consultations with different startups. These consultations were conducted between 2020 and 2022. As explained in \ref{ap_HITLlit}, we further substantiated the taxonomy by comparing our initial observations with the existing literature.

\section*{Statement on use of Large Language Models}

We have used Gemini Pro 2.5 Deep Research to help with literature search, and Gemini Pro to help with grammar and formulation in selected places.

\newpage

\bibliography{iclr2026_conference}
\bibliographystyle{iclr2026_conference}

\newpage

\appendix

\section{Introductory Material}

The main purpose of this appendix section is to provide further details on some of the terminology we use that we did not have space for in the main part of the manuscript, which we now signpost here for convenience.

In \ref{ap_harms} we specify the interpretation of \textit{harm} that we make use of in this manuscript, as well as the associated terms \textit{risk} and \textit{safety}, connecting these to existing literature. Next, \ref{ap_existing terms} further expands on the many terms related to HITL setups that appear in the literature, with references for each. And  finally, in \ref{ap_existing concepts} we discuss how existing conceptions of HITL are presented in the literature, how these relate to our work, and what we plan to do in the rest of the manuscript to extend and formalise these concepts.

\subsection{What we mean by harm, risk, and safety}\label{ap_harms}

Broadly speaking, we take `harm' to mean the infliction of some form of damage, be it physical or otherwise. Related to this, `risk' refers to the chance of some harm(s) occurring. And `safety' is the reduction or mitigation of risk, and hence avoidance of harm.

We use these terms in the broadest possible sense when talking about the failures of HITL setups. For example, strictly speaking, the (human) driver of a car falling asleep at the wheel is not a `harm', but rather a (normatively-defined) `wrong'~\citep{diberardino2024algorithmic}. However, it may rapidly lead to a (rather severe) harm, such as a car crash. So when we talk about harms, we also include these `wrongs'; events and actions which, though not necessarily harms in their own right, would almost certainly lead to harms in the strict sense of the word. And the same goes for how `harm' is interpreted in the definitions of risk and safety.

Substantial further work has studied the specific actions within sociotechnical systems that lead to material harm~\citep{diberardino2024algorithmic}, as well as what these harms can potentially be~\citep{shelby2023sociotechnical} and how one may take steps to actively avoid them~\citep{dobbe2022system}. Indeed, it may be insightful to relate these notions to failure modes (\S \ref{Failure modes}) and HITL setup structure (\S \ref{reductions}) from this manuscript as part of future work.

\subsection{Existing terminology}\label{ap_existing terms}

The existing terminology describing human involvement with ADMSs varies with differing degrees of human interaction and control. HITL setups have humans actively integrated into the ADMS's operational cycle (here we exclude human input in the training or development phase), while Human-on-the-loop (HOTL) have humans primarily acting as supervisors who intervene only when necessary~\citep{nothwang2016human}. Setups where the ADMS is designed to operate without direct human input or intervention can be termed Human-out-of-the-loop (HOOTL)~\citep{wagner2011taking}. A higher level perspective is given by Human-in-Command (HIC), whereby humans determine the high-level functioning of the ADMS~\citep{anderson2022human, kowald2024establishing}. To bridge the legal and ethical responsibility gaps~\citep{matthias2004responsibility} that can exist in such setups, the concept of Meaningful Human Control (MHC) was introduced to study how much influence a human should have on the execution of an action and what necessary levels of cognitive and moral awareness they should possess~\citep{davidovic2023purpose, roff2016meaningful, Abbink.2024}. In this context, \citet{Green.2022} speaks of three human oversight policies: 1) `restricting solely automated decisions', 2) `emphasising human discretion', 3) `requiring meaningful human input.' Later, in \S \ref{reductions}, we show how existing HITL terminology can be understood from the perspective of formal computability theory, and use this to deconstruct (3) above into a more fine-grained categorisation.

\subsection{Existing conceptions of HITL}\label{ap_existing concepts}

HITL setups can in certain situations mimic the trolley problem (i.e., the human needing to choose between multiple undesirable outcomes), particularly when the human has no ability to shut off the ADMS entirely, but often also go beyond it in legal, moral and technical complexity (cf.~\citet{Steenson.2021}). Recent research suggests that while HITL setups lead to more uptake of ADMSs, they potentially also decrease accuracy~\citep{sele2024putting}. \citet{Green.2022} particularly discusses the empirical evidence for human oversight policies of ADMSs, arguing that these generally fail to address the fundamental flaws in controversial government algorithms while simultaneously offering legitimisation of the algorithms and the protection of vendors and agencies from accountability. \citet{Elish.2019} similarly raises the point that HITL setups can end up protecting the ADMS rather than the human, speaking of humans as the `moral crumple zone' taking on accountability and enabling potentially faulty ADMSs to stay in place. Overall, this literature suggests that a closer scrutiny of HITL setups and their failure modes is necessary. 

The lack of a well-defined meaning of what a HITL setup can and does involve, and why such setups may be beneficial in sociotechnical systems, is a recognised problem. Recent literature has attempted to specify potentially desirable HITL setups in medical contexts. \citet{co-reasoners} have argued that both clinicians and patients should be included as `co-reasoners' in a medical ADMS context, making their own judgments about if, how, and why to use this technology, as well as how to use its results. Accordingly, they argue that such a HITL setup is valuable in that it forces questioning by both parties about whether to use ADMS technology thereby justifying its use, reduces automation bias by encouraging doctors to not rely on an ADMS without justifying its use to the patient, and uses an ADMS not just as a tool to generate answers but as `discussion' prompts based on the values and aims of both patients and doctors.  Building off the setup in \citet{co-reasoners}, \citet{griffen2024human} have argued for a kind of proliferation of HITL setups in medical ADMSs, highlighting the potential role of other clinical staff, and the value of having a diverse range of humans in these setups.

Recognising that human experts must play an active role in HITL setups, \citet{natarajan2025human} argue to shift the perspective and call such setups AI-in-the Loop (AI$^2$L), whereby it is the ADMS  that is part of a larger sociotechnical (human-led) process. Recognising the different degrees to which humans are, and should be, involved in producing functioning HITL setups, we present a classification of these in \S \ref{HITL reductions} (using Turing and many-one reductions) to concretise the language for further legal and ethical analysis. 

HITL is frequently presented as a critical safety measure in high-risk domains~\citep{EUAIAct.12.07.2024}, such as autonomous driving~\citep{Huang_2024} and the military~\citep{HITL_in_military}. However, the scalability of human oversight has been increasingly questioned (cf.~\citet{chiodo2024regulating}), especially for advanced ADMSs. The challenge of \textit{scalable oversight} highlights fundamental limitations: human supervisors may struggle to meaningfully oversee ADMSs whose cognitive capabilities surpass their own across multiple domains~\citep{amodei2016concreteproblemsaisafety, bowman2022measuringprogressscalableoversight}. Our work in this manuscript, on both the formal classification of HITL setups, and their failure modes, will help shed light on the positive and negative safety aspects of different HITL setups.

\newpage

\section{Computational Reductions}

The main purpose of this appendix section is to elucidate some additional background and insights from reductions that we did not have space for in the main part of the manuscript, which we now signpost here for convenience.

In \ref{ap_human-nondet} we give full details of what we mean by conceiving of the human in a HITL setup as an oracle. This includes a thorough treatment of the technical and formal machinery and background used, as well as certain clarifications on what actually happens in these setups in practice. Following on from this, \ref{ap_ping-pong} gives a diagrammatic illustration of the three HITL setups we define.

\ref{ap_intermediate} discusses intermediate setup types between \many{} and \turing{}, and why we have chosen to avoid covering them explicitly in our analysis. \ref{ap_routes} gives an expanded description of our route-planning example demonstrating our three setup types.

We then proceed to cover the computational strengths (\ref{ap_involvedbest}) and socio-technical benefits (\ref{ap_involvedbenefits}) of \turing{}s, and in \ref{ap_black box} give a diagrammatic representation of how they `open the black box' of the computation for additional scrutiny.

We finish by discussing the difficulties of determining setup types in practice (\ref{ap_HITLtypes}), with specific reference to setups which might appear to be one type, but in practice can be mimicked by a less-powerful one (\ref{ap_boundedtruth}).

\subsection{Conceiving humans as oracles}\label{ap_human-nondet}

In our formalisation of the notion of human-in-the-loop we represent the human by a fixed oracle function $f$ which is used in some oracle machine $T^\bullet$. This may suggest that we are assuming human decisions are transparent, deterministic, and truthful (perhaps even requiring some kind of omniscience). On the contrary, our view of human--machine setups as oracle machines implies quite the opposite.

The operation performed by the human in a HITL setup is precisely one which cannot be automated. Practically speaking, this is the reason the human enters the computation in the first place. The reasons for this difficulty in automation can be manifold, depending on the nature of the operation. It may be difficult for machines to do or even grasp, and involve questions whose answers are subjective, require life experience to give, rely on some sort of `moral judgement', or depend on unforeseeable circumstances. Often the human may be confronted with questions that do not admit a `correct' answer in the same way evaluating some arithmetic expression does, e.g., if they entail judgements on values, morality, or emotional response. In regards to subjectivity and morality, some fundamental computational problems are discussed in detail in \citet{MooralityComp}, and more recently \citet{EthicalWeapon} argued that at least (ethical) decisions cannot be made by systems which are predictable. Hence these problems are left for the human to grapple with; a practicality whose outcome (almost by definition) cannot be predicted by the machine developer. If the developer did know what the human would answer on some query, or how the human might `compute' such an answer, they could simply implement that response or computational process in the machine. This we formalise by saying that the developer implements the machine $T^\bullet$, which is essentially an algorithm that asks questions at certain times. Such questions are those which the developers cannot give an algorithmic answer for, and instead require a human to answer. The instance answering the questions is the oracle function $f$; the developer does not know the precise behaviour of such a function, but has to treat the outputs as computationally meaningful (see below). The human is to act as the oracle; they will answer the questions to the best of their ability (ideally), but of course may have a myriad of failure modes as discussed in \S \ref{Failure taxonomy}.

Once the decision-making procedure is put into action, the human will provide some inputs that are processed by $T^\bullet$. Therefore, in effect the human is providing some function $f$. Crucially, the developer has no influence on $f$ (or at best very weak influence) and hence the function is for all practical purposes fixed from their perspective---they have to design the system to make the best out of whatever the human will answer. Yet, this function need not be literally fixed in advance. The human could change their mind over the course of the decision-making process, and thus answer the same question differently if asked later.\footnote{This can be modelled by the function if the time at which the query is asked is part of the `input' the human receives.}  Indeed, it could be that in the morning the human flips a coin determining their moral compass for that day, or they are exhausted and behave differently than they would at other times,\footnote{To give just one example from a vast literature on physiological effects on decision-making: radiologists' judgements on prostrate imaging results become more pessimistic later in the day \citep{HungryJudge}.} or they have learned new things and improved their skills over time---many settings in which the human behaves in ways that are not transparent or even determined in advance are conceivable. It could even be that there are multiple humans who together provide the input, or different humans at different times or at different points in the human--machine interaction. Thus if we conceive of the human as an oracle, and say that any particular oracle is given by a function $f$, this does not impose any assumptions whatsoever on how humans behave. We also do not suggest in any way that the specific person to serve as the human in the HITL setup needs to be known in advance when we say that $f$ is for practical purposes to be viewed as fixed.

The distinction we make between the properties of the oracle function and the computational power entailed by the oracle machine mirrors that encountered with random machines.\footnote{See \citet[p.~33]{melkebeek} for some introductory discussion of random Turing machines; there the oracle is just a string which is read bit by bit, but the point stands.} In a random machine, any particular computation is guided by an oracle function $f$ which is a bona fide (deterministic) function. However, the point of random machines is that the oracle function is drawn from a probability distribution which then induces a probability distribution on machines (hence, computed functions). In this sense, `the oracle' is random, because the probability distribution on $f$ is what introduces randomness into the computation. This is even though any particular oracle \emph{function} $f$ itself is deterministic, and some of these may produce the correct output while others do not. A random machine obtains its strength through the fact that it is designed to call a random oracle, for which the properties of any individual oracle function are irrelevant (indeed, $f$ could be \emph{any} function from words to words); it only matters that \emph{most} oracle functions yield the correct output. Likewise, a decision-making procedure involving a HITL setup obtains its strength through sufficient use of the human, but it does so precisely because the behaviour of the human is not predictable. What is important for us is not so much what function $f$ is `computed' by a human oracle, but rather how any reasonable such function could fit into the decision-making procedure modelled by $T^\bullet$. 

Along the same lines, we would like to make a few clarifications.

\begin{enumerate}
\item In the computation, the machine $T^\bullet$ has to use whatever answer the human provides, treating it as `true' to some extent. This is a trivial observation: If the human input was treated as completely untrustworthy, it could not be used to generate any insight. Thus a developer has to allow the oracle to answer questions `however it sees fit' and design the system to process any reasonable answer as `serious'. This of course does not mean that the (human) oracle is actually truthful or `correct', or is even providing the `best possible' response. Indeed, developers should take into account that it may not be.
\item  As we point out in \S \ref{legal-moral}, there may also be legal or moral requirements for a human to evaluate and respond to a question arising in the decision-making process, even if it seems a machine could answer it. In this case, the contribution of the human is moral, rather than strictly computational, as phrased above. Still, the human plays an integral role if the overall decision by the human--machine system needs to reach some legal or moral status, and so the human is needed to produce the output.
\item We do not mean to imply that a literal Turing machine which operates in the way we describe would be a \emph{practical} model of modern computational systems. Rather, our perspective is that it seems productive to transfer notions from computability theory to human--machine systems. Our formalisation is supposed to enable this conceptual move.
\end{enumerate}

To summarise, the point of describing HITL setups in terms of the formalism of oracle machines is not to view (or idealise) humans as oracle functions, but rather to view human--machine decision-making setups as oracle machines.

\subsection{Diagrams of how our HITL setups operate}\label{ap_ping-pong}

\begin{figure}
 \centering
\includegraphics[width=0.9\textwidth]{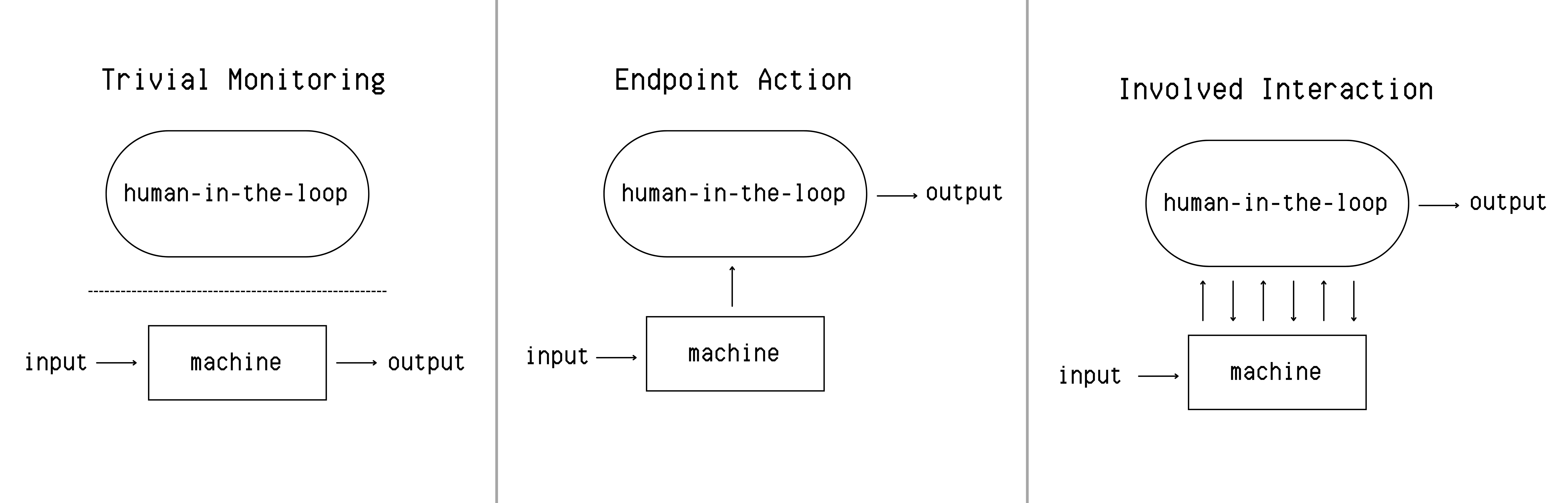}
    \caption{Operational diagram of each of our HITL setups.}
    \label{fig:setups1}
\end{figure}

In Figure \ref{fig:setups1} we illustrate the operation of our HITL setups: \trivial{},  \many{}, and \turing{}. Note the computational `ping-pong' between the machine and the human in the \turing{} setup, which significantly increases human input into the process.

\subsection{Intermediate HITL setups}\label{ap_intermediate}

Why have we omitted discussion of restricted cases of \turing{}, such as where the machine can only ask the human oracle a bounded finite number of real queries (corresponding to a Turing reduction where the oracle can be consulted only a bounded number of times)? Computationally, this would lie properly between a many-one reduction and a `full' Turing reduction. This omission is deliberate, as it would invite discussion also of the many other intermediate reduction types between many-one and Turing\footnote{Including bounded truth-table reductions, truth-table reductions, and weak truth-table / bounded Turing reductions (the last of these we discuss in further detail in \ref{ap_boundedtruth}). See~\citet[p.~83]{soare87} or~\citet[Section 2.4]{melkebeek} for details.}. Conceptually, an \many{} describes the `simplest' way a human can interact with a machine in a computationally meaningful way, and an \turing{} describes the `most complicated'; these two types alone give us a deep and rich area to analyse. It may well prove valuable to replicate the technical, legal, and moral analysis done in this manuscript on some of these intermediate reductions (e.g., when the machine can do its calculation, hand over to the human to do some more, then take back the human output and do some final machine calculation; having `one round of ping-pong'). The technical, legal, and moral possibilities may well be endless, so we have elected to examine the extremal cases for now: \many{}s, and \turing{}s, which correspond to many-one reductions and Turing reductions respectively. 

The key notion for our purposes is thus that of the real query. Some care is needed when applying this notion in practice. To illustrate this very simply, consider a machine that attempts to guess an integer $n$, where it is known that $n = m + k$ for some integers $m, k$ about which nothing is known. The machine might ask whether $m$ is even and then whether $k$ is even. According to our definition, the first question is not a real query because by itself it does not change the set of possible outputs (any integer can be written as an even integer $m$ plus some integer $k$). The second question then is a real query, as it determines for instance the parity of $n$. However, if the first question was somehow omitted, the second question would not be a real query for the same reason the first is not. This simple example shows that whether something is a real query depends not just on the question but on the whole computation. These interactions between questions may be complicated in some systems, but the key point remains: A query is only substantial insofar it leads to a real query.

A related question which arises from our analogy to computability theory  is what resource-limitations on the machines and oracles in question play a role. In the theory of computation quite substantial attention was paid to the question of what happens if oracle machines which run in polynomial time are given oracles for various hard but computationally feasible problems. Such is the theory of complexity classes from and around the polynomial hierarchy, such as $\mathsf{P}$, $\mathsf{NP}$, $\mathsf{BPP}$, $\Pi^{\mathsf{P}}_2$, or $\mathsf{PSpace}$ (see \citet[section 2]{melkebeek} for definitions of these classes and (ibid. Section 2.4.3) for complete problems). Mathematically this theory is quite different from the theory of reductions without resource-limitations (i.e., classical recursion theory). To some extent our taxonomy of failure modes (\S \ref{Failure taxonomy}) might be viewed as a comment on the resource limitations of human oracles. It may be useful to explore such viewpoints in later work, but we refrain from doing so as it does not seem to clarify our basic point in this manuscript.

\subsection{Route planning machines demonstrating different HITL setups}\label{ap_routes}

Consider a route-planning machine in a HITL setup with the human driver of a car. It may demonstrate any one of the following HITL setups:

\textit{\Trivial{}:} The human could enter the origin and destination, and the machine could then present a driving route. The human then has the choice to take the route, or not. Here, the human has no input to the computation process, and can only `turn off' the machine (by ignoring its output).

\textit{\Many{}:} The human could enter the origin and destination, and the machine could present the human with several different options, perhaps labelling them as `fastest', `most fuel efficient', `most reliable', `passes fuel stations regularly'.  The human can then choose between these. Here, the human takes over at the end of the machine's work to finish the computation and produce the route to be taken (from the list presented to them).

\textit{\Turing{}:} The human could instead input into the machine `I want to visit my sibling'. The machine could then start computing, and asking the human a series of questions. It might start with `when are you travelling?', take the answer, and do some more computation before coming back with `when do you need to be there by?', then take that answer and do a bit more computation and come back with `what do you need to bring?', then `is anyone else travelling with you?', and so on,  each time doing some computation between each question. These questions cannot be stacked all at the start, as the answer to earlier questions might determine which later questions are asked (and the number of questions may not be a priori bounded at the onset). The machine \textit{may} then present some sort of optimised driving route. Or it may produce a very different output, such as suggesting to take the train as there is bad traffic, or to go on another day  closer to the sibling's birthday. Or it may even advise against travel due to adverse weather. Here, the human has regular, meaningful, and not a priori bounded input into the computation.

\subsection{The strength of \turing{}s}\label{ap_involvedbest}

Computationally, a Turing reduction between functions is viewed as weaker than a many-one reduction, as many-one implies Turing. This comes from a viewpoint that \textit{fixes a computational problem}, and then asks what the space of oracles it reduces to is. So a Turing reduction is considered weak, because the problem reduces to more oracles; a many-one reduction would reduce it to fewer oracles so the reduction itself is strong. Thus, if one tries to many-one reduce a given problem to some oracle rather than Turing-reduce it, then one (generally) requires a stronger oracle, which in the context of HITL setups means a more competent human. Therefore, our concern is the opposite: we advocate maximising the space of functions that can be computed with a \textit{fixed oracle} (a human). With that perspective, a Turing reduction is considered strong, because more problems Turing-reduce to the given oracle than if we used many-one reductions instead. In a HITL setup, the oracle is fixed (the human), so limiting the setups to \many{}s limits the space of problems that can be solved. Therefore, making the space of problems solvable as large as possible is achieved by maximising the influence of the human, i.e., by using an \turing{} setup.

\subsection{The benefits of using \turing{}s}\label{ap_involvedbenefits}

What exactly are the benefits of an \turing{} HITL setup? In short: because human (oracle) influence can add valuable, desirable insights throughout the overall computation, in particular at times where it is most needed by the machine.

Firstly, the human has more influence on the overall computational process, and thus on the outcome. With more, potentially unbounded, real human queries, the \textit{agency} of the human is increased. Secondly, and related to this, the human has more opportunity to input their judgements and values into the overall computational process, thus giving more opportunity for the human to ensure the machine computation is \textit{aligned} with its values. Thirdly, by intervening more often, the human has more opportunity to identify (and rectify) problems and safety issues within the machine computational process, before a final output or behaviour occurs. Increased human queries `bakes in' the potential for increased human scrutiny. This aids with the \textit{safety} of the HITL setup. Finally, and related to this, with more human queries, the machine is articulating intermediate steps that are human-interpretable more often (as a human input is sought), thus aiding with \textit{transparency} of the  computational process.

Overall, these four aspects come together to improve the \textit{reliability} of the HITL setup.

\subsection{Opening the black box via \turing{}s}\label{ap_black box}

\begin{figure}
 \centering
    \includegraphics[width=1\textwidth]{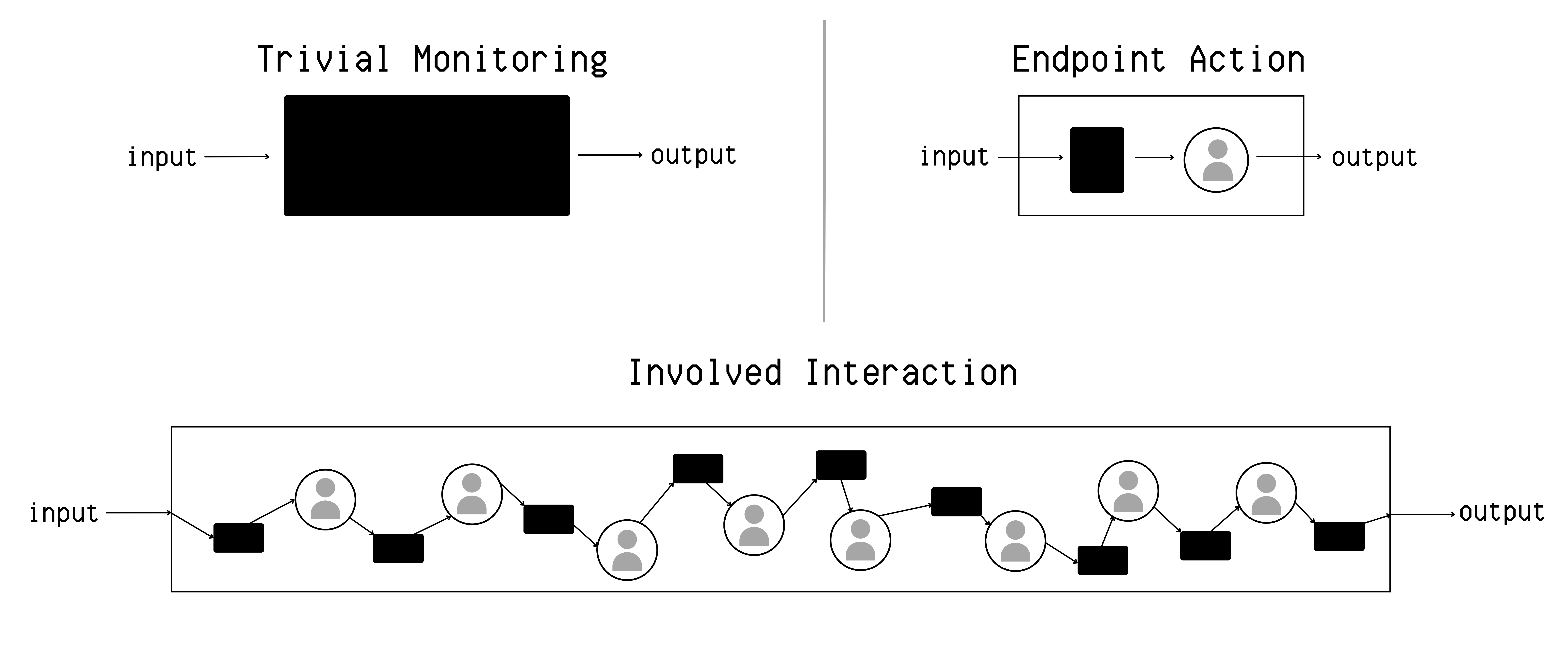}
    \caption{How real queries in HITL setups can unmask black-box computation.}
    \label{fig:setups2}
\end{figure}

In Figure \ref{fig:setups2} we illustrate the gradual `unmasking of the black box' as one increases human involvement, from \trivial{}, to \many{}, and \turing{}. Note the chain of computation in the illustration of the \turing{} setup; this is a different way to view the computational `ping-pong' between the machine and the human, and shows how this ping-pong serves to help unmask the black-box computation by the repeated handovers between the machine and the human. The human activity, and human--machine handovers, are all very interpretable to an observer.

This is because each time the machine needs to `ask a question' to the human, it must precipitate its inner computation in a human-interpretable way, and then take a (human-interpretable) input. As  more such questions that are asked, the overall box becomes proportionately more filled with these human interpretable questions-answers, and so the black-box steps make up less of the process. As the number of such questions becomes very large, the remaining `black box' parts of the computation become proportionately quite small, allowing humans to get a very good overall understanding of the  computation (even if they cannot see every minute detail). As shown in Figure \ref{fig:setups2}, the \trivial{} setup is completely black-box, the \many{} setup has a large black box process (but significant human-interpretable process), and the \turing{} setup is mostly human-interpretable steps (with very small black-box processes between them).

\subsection{Determining HITL types}\label{ap_HITLtypes}

The level of formalism for HITL setups presented in \S \ref{HITL reductions} is an important aspect of our manuscript. It is heavily relied upon later in \S \ref{legal-moral}, giving ways to prevent developers from circumventing the regulatory \textit{intentions} behind any informal definitions of what a HITL setup should be. In particular, our definition of a real query, and the associated formalism around how each HITL setup type makes use of real queries, prevents such setups from having a `veneer' of significant human involvement. This formalisation was the best way we could find to properly describe these setups in a way that could not be easily `cheated' or done `in name only' without implementing the meaningful effect intended. However, determining HITL setup types, and what they achieve, is a challenging task from each of a technical, legal, and moral perspective.

From a technical perspective, showing the non-existence of a reduction is difficult. While it is fairly straightforward to demonstrate some setup constitutes, say, a many-one reduction, by simply inspecting the structure of the reduction, that would not (in our terminology) show that a HITL setup is an \many{}. To achieve that, one would also need to show that the machine was not total, and that it always asks a real human query during the computation. Similarly, to show a HITL setup is an \turing{}, one would need to show that it is a Turing reduction, but also show that the machine always asks more than one real human query during the computation. Heuristically, while it is `easy' to show that there is a human in the loop (as this is simply done by saying that a  human is present), it is much harder to (computationally) verify how much the human will actually be asked to do. 

From a legal perspective, showing that the human will do something meaningful is also difficult. One could have a HITL setup where there are real human queries which are completely deterministic and replicable by the  machine (i.e., asking the human `What is 1+1?'). Asking the human a series of `pointless' questions (that the machine could just as easily carry out itself) may violate the legal principles of `meaningful' or `effective' human input, as discussed in \S \ref{legal-moral}. 

From a moral perspective, showing the HITL setup is not simply a facade masking a less-involved process is also difficult. One could have a HITL setup where there are real human queries which the machine cannot do itself (i.e., bringing in genuine human judgement), but then the remainder of the (machine) computation simply ignores the input, or worse still, actively inverts or goes against the input. Asking the human a question with a (seemingly) important answer, and then not using that, would at best be a terrible oversight by the machine developers, and at worst be completely disingenuous and unethical of them. If the machine were to actively go against a human input, and that had been a deliberate design feature, then that would be completely immoral.

\subsection{Bounded truth table reduction}\label{ap_boundedtruth}

In the \many{} scenario of our  route-planning example in \S \ref{HITL reductions} and \ref{ap_routes}, the machine could have instead asked the human a fixed finite series of questions such as `Do you prioritise speed or  efficiency?', `What is your maximum acceptable distance between fuel stations?', etc., and done computation between each, to find a single route to then present. This is an \turing{}, but in a precise sense it is not `fully using the potential" of a Turing reduction. Namely, this process corresponds to a type of intermediate reduction as mentioned in \ref{ap_intermediate}: a bounded truth table reduction. The crucial difference is that the uses of the oracle tape are bounded independent of the oracle. This process can be viewed as one that presents only one query and performs only a trivial operation after this. Namely, an oracle machine that asks a (human) oracle a (predetermined, finite) set of questions `in series' one after the other, is computationally equivalent to one which does a slightly different computation with these options `in parallel' and presents the conglomeration of questions to the oracle at the very end, phrased as one (big) question. So essentially these setups are the same as those which have only one real query, which are slightly stronger\footnote{The Halting Problem does not many-one reduce to its complement, but there is a Turing reduction with only one query. It remains to be discussed when this distinction matters for human oracles.} than those for \many{}, but weaker than those in \turing{}. Not all Turing reductions are made equal, and even among those which are not many-one reductions, some may behave differently than others from the perspective we advocate here.

\newpage

\section{HITL Failure Modes}

The main purpose of this appendix section is to motivate and expand upon our taxonomy of HITL failure modes, with more detail than would otherwise fit in the main part of the manuscript. We now signpost this here for convenience.

The main part of the appendix, \ref{ap_HITLlit}, explains the origins and motivation of our taxonomy of failure modes, based in part on our experiences in industry. This is supplemented by extensive existing literature on the topic which supports our taxonomy. We also include a much-elaborated table of failure modes for our five failure categories.

To round off the appendix we give two case studies, covering additional examples of failed HITL setups that have occurred. The first of these (\ref{ap_MCG}) is covered in general terms. The second (\ref{ap_Notre}) is dissected and analysed according to our taxonomy from \S \ref{Failure taxonomy}, as was done for the case study in \S \ref{Uber example}.

\subsection{Literature on HITL failure modes}\label{ap_HITLlit}

The empirical foundation of the taxonomy given in \S \ref{Failure taxonomy} is primarily derived from a series of ethics consultations conducted between 2020 and 2022. During this period, one of the authors served on the ethics advisory board for an AI startup accelerator program \textit{[anonymised for peer review]}. As part of this engagement, the author provided ethics consultations to over 25 AI startups and conducted intensive, in-depth ethics workshops with three additional companies.

These consultations provided a valuable opportunity to observe how AI practitioners conceptualise system failures. A recurring theme emerged: participants, who often possessed strong technical backgrounds in fields such as computer science, mathematics, and engineering, demonstrated a sophisticated understanding of potential technological failure modes. This technical expertise aligns directly with the first category in our proposed taxonomy: `Failure of the machine components'.

Conversely, the consultations revealed that the complexities arising within broader sociotechnical systems, particularly those with an (implicit) human-in-the-loop, were less consistently understood. The dialogues surrounding the limitations and potential failures of human--machine interactions highlighted critical areas for conceptual development, and showed that a more fine-grained taxonomy of failure modes is warranted than simply categorising all other failures as `human error'. These interactions, therefore, were crucial in helping us develop essential insights, thus allowing us to identify the subsequent categories of our taxonomy.

The issues that arose with the startups consistently coalesced around four foundational themes beyond the purely technical. These themes, which informed the taxonomy's structure, included:

\begin{itemize}
    \item Process and Workflow: For example, the design and strategic importance of robust data segregation practices, illustrated by questions concerning the rationale for utilising unique encryption keys for different clients to mitigate a (human) loss of a key.
    \item human--machine Interface: For example, the psychological and ergonomic significance of interface design choices, such as the colour and typography of warning messages, in conveying critical information.
    \item The Human Component: For example, the necessity of anticipating and modelling a wide spectrum of user behaviours, including those that are unconventional or counter-normative such as the unanticipated uploading of sensitive information by users to an insufficiently-secured cloud storage system.
    \item Exogenous Factors: For example, the complex ethical and legal considerations involved in the contractual allocation of liability between software providers and B2B end-users.
\end{itemize}

A common theme in these `mixed' failure modes was that startups, and in particular the technical developers themselves, did not necessarily see such failure modes as their areas of responsibility. Multiple times, the author heard variants of the statement `This problem would be human error, and not a problem in our technology, so we are not responsible for it'. The contrast between nuanced failure modes and the general idea of `being a user problem', helped spawn the idea that with poor sociotechnical choices at the design stage, the humans participating in the HITL setups were being (unintentionally) \textit{set up to fail}. 

The fineness/coarseness of the taxonomy as 5 failure categories is a pragmatic interpretation of the topics that came up during these interactions, and how the various developers and managers were able to deal with them. We tried to keep in mind what appears to be best-suited to those designing, developing, and deploying HITL setups, and we note that there may well be other, more philosophically-oriented breakdowns one might use. 

Many of the individual failure modes which were observed during these consultations can by now also be found in the wider literature, albeit there they are presented with a different structure and/or in unstructured ways. For example, the recent literature survey written by \citet{10.1145/3630106.3659051} develops a taxonomy for general human oversight in the context of AI. While they only consider the failure categories of technical design, human, and environment, they still discuss many of the individual failure modes presented in our taxonomy. We note that the differences between our taxonomy and that of \citet{10.1145/3630106.3659051} are a reflection of the empirical experience of conducting ethics consultations with startups, whereby discussions of the failure modes were naturally ordered by `increasing human-ness' or `lack of technicalities,' and many concerns emerged at the boundaries between human practices, abstract processes and technological details (thereby requiring us to have 5 categories). Moreover, \citet{chiodo2023manifestoresponsibledevelopmentmathematical} cover the harms and failure modes of general mathematical work in their `10 Pillars of responsible development', which also has close connections to our taxonomy and its breakdown into various failure modes.

Having studied failure modes for AI-supported governmental decisions, \citet{Green.2022} also notes that the introduction of a HITL setup can harm the overall safety of a sociotechnical system; something that was also observed during the ethics consultations, as the `this is not a technical problem' attitude was often accompanied by placing  a lot of potentially unwarranted hope on human oversight, thus creating a false sense of security in the AI product or service being produced.

Regarding the consultations with startups, the author also found that discussions on failure modes in automated industrial processes, as well as the commonly discussed `Swiss cheese model' \citep{larouzee2020good} for accidents, whereby an accident can happen despite many defensive layers if the `holes' of each layer are properly aligned, were still relevant in HITL contexts. For example, \citet{Bainbridge1983} wrote about the automation of industrial processes, \citet{endsley1995toward} studied the situational awareness of humans in automated processes, \citet{Sarter.1997} looked at automation surprises, and~\citet{reason1990human} for human errors---the latter two have influenced categories 3 and 4 of our taxonomy, focusing on the human and the human--machine interface. Understanding that many of these failure modes are not unique to AI systems or more general ADMSs, but can also be found in other (industrial) situations helped convey the spectrum of responsibilities to various startups.  We present here in Table 1 an extended (but still non-exhaustive) breakdown of our taxonomy, with additional failure modes included, extending the list from \S \ref{Failure taxonomy}. This is done in tabular form for ease of reading:

\newpage

\begin{center}
    Table 1: HITL failure modes
\end{center}

\begin{tabular}{|>{\raggedright\arraybackslash}p{2.33cm}|>{\raggedright\arraybackslash}p{2.33cm}|>{\raggedright\arraybackslash}p{2.47cm}|>{\raggedright\arraybackslash}p{2.33cm}|>{\raggedright\arraybackslash}p{2.33cm}|}
\hline
\textbf{Failure of the machine components} & \textbf{Failure of the process and workflow} & \textbf{Failure at the human--machine interface} & \textbf{Failure of the human component} & \textbf{Exogenous circumstances} \\ \hline
$\bullet$ Unexpected inputs or outputs \par $\bullet$ Problematic machine evolution or self-adaptation \par $\bullet$ Hallucinations \par $\bullet$ Reasoning errors \par $\bullet$ Overfitting of training data \par $\bullet$ Biased or other erroneous outputs \par $\bullet$ Unfalsifiable outputs \par $\bullet$ Lacking `common sense' \par $\bullet$ Morally unacceptable outputs \par $\bullet$ Other unexpected behaviour & $\bullet$ Insufficient power of the human \par $\bullet$ Insufficient self-control/ \par independence \par $\bullet$ Insufficient reaction time \par $\bullet$ Unrealistic expectations \par $\bullet$ Delayed notification \par $\bullet$  Lack of disaster planning \par $\bullet$  Insufficient management support \par $\bullet$ Insufficient psychological support \par $\bullet$ Lack of rest \par $\bullet$ Conflict of interest \par $\bullet$ Other process and workflow failures &  $\bullet$  Incomprehensible or incomplete outputs \par $\bullet$   Complex or poorly designed user interface \par $\bullet$ Constantly changing user interface \par $\bullet$  Insufficient training \par $\bullet$ Poor documentation \par $\bullet$  Transition failures between different humans \par $\bullet$ Other HCI adaptability failures \par $\bullet$  Other epistemic failures \par $\bullet$ Other interaction failures &  $\bullet$ Cognitive bias \par $\bullet$ Automation bias \par $\bullet$ Confirmation bias \par $\bullet$  Fatigue \par $\bullet$  Incongruous intentions \par $\bullet$ Stress or overload \par $\bullet$ Lacking courage \par $\bullet$ Lacking motivation \par $\bullet$ Lacking self-awareness \par $\bullet$ Lacking humility \par $\bullet$ Onset of groupthink \par $\bullet$ Other human-centric failures &  $\bullet$  Unreasonable laws \par $\bullet$  Unreasonable societal expectations \par $\bullet$ Conflicting requirements \par $\bullet$ Misaligned objectives \par $\bullet$ Political pressure \par $\bullet$ Unexpected exogenous shocks \par $\bullet$ Poor safety culture \par $\bullet$  Inappropriate workplace requirements \par $\bullet$ Insufficient resources \par $\bullet$ Other external pressures\\ \hline
\end{tabular}

\mbox{}

One can further see very recent literature connected to each specific failure category given above:

\begin{enumerate}
    \item Regarding failure of the machine components, \citet{barassi2024toward} investigates what the term `AI errors' even means given the new and complex world of LLMs and other multi-modal models. And \citet{kim2025beyond} discuss how ADMS errors can systematically feed into human performance and human errors further down the line.
    \item Regarding failure of the process and workflow, \citet{rosenthal2024michael} conduct a deep investigation showing many humans in HITL setups may simply be unable to detect \textit{systemic} bias in the overall output of an ADMS (in their study, a hiring algorithm).
    \item Regarding failure at the human--machine interface, \citet{tsai2025effect} have studied human performance on ADMS-assisted verification tasks, showing a substantial increase in cognitive processing and cognitive load on the humans when the ADMS assistance was activated due to the increase in content being displayed to them.
    \item Regarding failure of the human component, \citet{alon2023human}  demonstrate `selective adherence' of humans, which is `the propensity to adopt algorithmic advice selectively, when it matches pre-existing stereotypes about decision subjects' \citep[p.~154]{alon2023human}.
    \item Regarding exogenous circumstances, \citet{Laux_Ruschemeier_2025} provide a critique of the EU AI Act, and in particular how its `focus on providers does not adequately address design and context as causes of automation bias' \citep[p.~1]{Laux_Ruschemeier_2025}.
    
\end{enumerate}

\subsection{HITL as part of an AI security scanner setup}\label{ap_MCG}

As presented in~\citet{Robertsonetal.2025}, in April 2025 two men were alleged to have brought firearms through the ADMS-powered security scanning setup at the Melbourne Cricket Ground (MCG). This was an \many{} HITL setup, with  scanners running an ADMS used to flag attendees who \textit{might} be carrying weapons, and then manual secondary screening used to conduct a final, definitive inspection. Initially, the  blame was conjectured to have been `human error' in the secondary scanning process.  However, as mentioned in (ibid.), with a potentially high false-positive rate, secondary (human) scanners may have faced the combined challenges of operator fatigue (having to scan many attendees in a short space of time as they arrived for the beginning of a match), as well as complacency (having to scan countless attendees, \textit{none} of whom were actually carrying weapons). The human scanners may have become quite tired, and many have lost faith in the ADMS (the scanner) as it overwhelmingly returned false positives. Indeed, shortly after initial reporting, it was revealed that the alleged offenders did not in fact have a metal detecting wand waved over them, even though the ADMS flagged them for additional checks \citep{RyanEddie.2025}. Both men were eventually prosecuted and found guilty of bringing firearms into the MCG \citep{Cosoleto.2025}.

\subsection{HITL as part of semi-automated fire detection setups}\label{ap_Notre}

As presented in~\citet{BennholdGlanz.07.05.2025, Peltier.18.06.2019}, in April 2019 a catastrophic fire broke out in the attic space of Notre-Dame Cathedral. The cathedral had an ADMS for fire monitoring with an \many{} HITL setup. When a fire was detected by the machine, a short message `fire' was sent to the monitoring office in the cathedral, without specifying exactly where the fire was (incomplete outputs). As is standard across France, fire alarms never automatically notify the fire brigade, so as to avoid false callouts (unreasonable laws). The employee on duty was only on his third day of the job and working unsupervised (insufficient training, unrealistic expectations). The employee was required to phone a guard, who  then had to physically check the attic; a 6 minute journey up many stairs (insufficient reaction time). Unfortunately, the guard got lost, and went up the wrong staircase to the attic of the sacristy, which was the adjacent building (insufficient training). The employee then called his manager rather than the fire brigade (lacking courage), but could not reach him (insufficient support, insufficient power of the human), and it took time for the manager to call back and instruct the guard to leave the sacristy, go down the stairs, and then climb another staircase to the attic of the cathedral. By the time the guard reached the attic of the cathedral, the fire was raging. The fire brigade was then called, over 30 minutes after the first detection of a fire. This was a fairly simple system, computationally speaking, but  nonetheless was a HITL setup that failed completely with spectacular consequences.

\newpage

\section{HITL in the Law}

The main purpose of this appendix section is to provide additional details  for the legal cases and arguments that we present in the manuscript, the details of which are far too long to fit in the main part of the manuscript. We now signpost this here for convenience.

We start in \ref{ap_SCHUFA} by covering the legal concept of automated decision making, and how this relates to our definition of \trivial{}. There, we detail the SCHUFA case, where the courts found that the the credit scoring company SCHUFA violated the GDPR by (inadvertently) carrying out automated decision making by adopting a \trivial{} setup.

We then highlight the legal benefits of \many{}s in \ref{ap_safeguarding duties}, showing that such HITL setups can help the human to carry out their legal and moral safeguarding duties to a better extent. We also show that SCHUFA could have averted violating the GDPR had it made use of an \turing{} setup.

In \ref{ap_Explainability} we detail how causation can be difficult to determine in an \turing{} setup. And in \ref{ap_Mesothelioma} we present the outcome of some legal cases related to mesothelioma where a principle of joint liability was used by the courts to bridge responsibility gaps, and how this might apply to \turing{}s when trying to establish legal liability.

\subsection{\trivial{} as automated decisions (GDPR)}\label{ap_SCHUFA}

At present, Article 22 of the GDPR generally prohibits \trivial{} and classifies it as a solely automated decision despite the presence of an active HITL setup. The GDPR requires that any ADMS used to make decisions with legal (or similar) effects need to reflect at least an \many{} HITL setup to break automation. The scope of Article 22 was affirmed by the Court of Justice of the European Union (CJEU) in the SCHUFA case~\citep{EuropeanCourtofJustice.07.12.2023}. Prior to SCHUFA, it was thought that if there was \textit{any} active HITL setup then this could not be a solely automated decision within the scope of Article 22. However, in the case, the judge opined that a HITL setup \textit{can} still exist within a solely automated decision if the human is just carrying out \trivial{}. This is because in \trivial{}, the human is not meaningfully influencing the decision-making process, and so is unlikely to have any impact on the ultimate output of the system, meaning that the decision-making is basically still automated.

The SCHUFA case concerned a credit scorer (SCHUFA) who created credit repayment probability scores through automated processing. SCHUFA then shared these scores with lender banks who determined whether they would lend money to individuals, using those scores. The issue in the case was whether an automated decision occurs between SCHUFA and the individuals, even though the bank lenders were in the middle of the decision making process (acting as `the human' in this HITL setup). However, since the bank lenders relied so heavily on the scores, in reality, it meant that they were simply applying the output of SCHUFA's decision. Thus, the CJEU determined that this was an example of a solely automated decision between SCHUFA and the individuals seeking a loan, showing that \trivial{} setups are generally considered to be automated if the human has no meaningful influence on the decision.

This case marked a significant shift in the law's approach to HITL by recognising that tokenistic humans cannot be used to enable companies to fall outside the scope of Article 22 and thus avoid the onerous obligations associated with this provision. If data controllers want to perform an automated decision with no human intervention or if the human is just carrying out \trivial{}, to be lawful, they must also have explicit consent from the decision subject (or another specific justification) that permits decision making with such limited oversight. Additionally, when performing automated decision making, the data controller must also have certain safeguards in place, such as the ability for individuals to request human intervention, receive explanations of decisions, and contest the outputs. 

In order for a decision making process to fall outside the scope of Article 22 and `break automation', the human needs to actively influence the decision making process. The CJEU provided little guidance on what the ideal HITL setup looks like since this was beyond the scope of the case. However, guidance from the UK's data protection authority~\citep{ICO2025AutomatedDecisionMaking} suggests that to avoid making an automated decision, data controllers need to ensure that the human weighs up the output and interprets it before applying it to the decision subject. This reflects \many{}, since the machine asks one real query and the human considers the machine's output and adds some more insight to make the final decision. 

Therefore, the law recognises \trivial{} as solely automated and generally prohibited (unless consent or other justification and safeguards are present). In addition, regulatory guidance suggests that \many{} is required to move data controllers outside the scope of automated decision making in Article 22. However, as shown in this manuscript, there are significant yet distinct ethical and computational implications between \trivial{}, \many{}, and \turing{} setups that the law does not acknowledge. 

We argue in this manuscript that \turing{} provides legal benefits that align with the broader responsibilities set out in the UK GDPR and EU AI Act. Therefore, building upon the jurisprudence of the SCHUFA case, data protection authorities should provide guidance to companies to implement at least \many{}, and preferably \turing{} as an ideal HITL setup to fall outside the scope of Article 22. Existing guidance lacks practical tools that ADMS developers can use to enable \textit{effective} and \textit{meaningful} oversight. As such, including the specific computational setups like \trivial{}, \many{} and \turing{} with their applicable legal obligations would help clarify the scope of automated decisions to technical audiences and provide developers with concrete methods to configure their human--machine setups accordingly.

\subsection{The legal benefits of \turing{}}\label{ap_safeguarding duties}

Article 14 of the EU AI Act states that `Human oversight shall aim to prevent or minimise the risks to health, safety or fundamental rights'. Similarly, Article 22(3) of the GDPR stipulates that data controllers implement `suitable measures to safeguard the data subject’s rights and freedoms and legitimate interests'. These safeguarding duties are onerous and it is difficult to understand how the human in a \trivial{} or \many{} setup would be able to minimise risks to an individual's human rights or safeguard a decision subject's legitimate interests. In \trivial{} and \many{} setups, the human may have no influence on, or understanding of, the computational process involved in the decision. 

Indeed, in these setups the human may face a completely `black box output'. For example, in an \many{}, the human may just receive a probability score for credit risk. Using only this, the human would need to determine whether it should be applied to an individual and also whether it infringes any values. Thus, \many{} and \trivial{} setups contain an inherent opaqueness which prevents the human from understanding the significance of certain inputs or factors on the output of the ADMS. As a result, the human may lack sufficient knowledge and ability to influence, effectively evaluate, and implement the decisions made by an ADMS to uphold their safeguarding duties. 

In \S \ref{moving}, we argued that in systems with weaker reductions such as \many{}, the human is not enabled to effectively influence the system, and thereby meet their obligations under GDPR or the EU AI Act. We therefore propose that \turing{}s are included in regulatory guidance to guide developers on how the human can meaningfully impact the system, thereby meeting legal requirements.

Carrying forward the example of the SCHUFA case (\ref{ap_SCHUFA}), if the lender was engaged in an \turing{} setup, the ADMS could query the human in the course of the computation which would enhance the human's understanding of whether the decision subject's rights or legitimate interests are being eroded. For instance, the human could be asked questions by the ADMS such as `is the individual's postcode relevant to the credit score?' or `what information should be used to substitute the lack of credit history?'. 

Additionally, \turing{} setups give the human a chance to watch and scrutinise more of the computational process, as mentioned earlier. This may help overall safety, as they may `spot a problem' early on, such as the ADMS asking questions about a credit applicant that should not be used as part of the assessment. Therefore, \turing{} avoids the `sloppy'~\citep[p.~434]{Crootof.2023} implementation of a human into human--machine setups, by enabling the human to better fulfil their safeguarding duties by evaluating the output of an ADMS in accordance with the rights and freedoms of decision subjects. 

Also, an \turing{} setup could have actually \textit{averted} SCHUFA's legal issues regarding the court's determination of their system as automated decision making (prohibited by Article 22). In that case, it was deemed that even though the intermediary bank \textit{appeared} to be acting as `the human' in an \many{} HITL setup by taking the credit score from SCHUFA and using it within their own processes for determining credit worthiness, the courts determined that in actual fact that the human was in a \trivial{} setup as they were merely `passing on' the credit score as a credit worthiness determination. What we see here is that in this ostensibly \many{} setup, the human `slipped back' to a \trivial{} setup by completely deferring to the machine output; a clear case of (a particular form of) automation bias in our taxonomy of HITL failure modes (see \S \ref{Failure taxonomy}); one which was deliberate and designed into the system which unintentionally resulted in serious legal implications. Had the human (the bank) properly maintained its (\many{}) role of using the machine output to feed into a human decision, then Article 22 of the GDPR would not have been violated by SCHUFA. 

However, as explained in \S \ref{Failure modes}, HITL setups naturally fail, for a wide variety of reasons (automation bias being one of them), and this should have been anticipated by SCHUFA. So what could SCHUFA have done to avert violating Article 22? One way would have been to properly implement an \turing{} setup, needing several interventions from the human in the machine computation. This would have ensured the decisions were not automated, thus avoiding a violation of Article 22. With the \many{} setup SCHUFA had in place, the human could simply `nod through' the machine output in an algorithmic way, by accepting all applications scored above a certain threshold, and rejecting all those below it, meaning the setup was given by a total function. As such, this broke the requirement of an \many{}, which is that `the oracle machine many-one reduces the computation to the human, but it \textit{does not} define a total function' (\S \ref{HITL reductions}). 

If, on the other hand, the setup was as \turing{}, the human could not have nodded through the machine queries throughout if they were indeed real queries (as defined in \S \ref{reductions}), as they may well have faced a query not admitting a yes/no answer. Of course, the queries might all be algorithmically solvable (and so the setup again reduces to an \turing{}), but here SCHUFA would have been \textit{incentivised} to ensure this did not happen, to avoid violating Article 22. Thus, SCHUFA would have been in a much better position to avoid carrying out automated decisions (i.e., avoid violating Article 22) had they implemented an \turing{} setup whereby the human (the bank) needed to answer questions throughout the computation process. By implementing an \turing{} (\S \ref{HITL reductions}) to break the automation bias of the human (\S \ref{Failure taxonomy}), SCHUFA could have prevented the slip back to \trivial{} and thus avoided violating Article 22 (\S \ref{HITL law}).

This demonstrates a real example where our arguments from \S \ref{moving} play out, illustrating the benefits of having a HITL setup with stronger reductions, and thus more human involvement and input.

\subsection{Why legal causation is difficult in \turing{}}\label{ap_Explainability}

\Turing{}s have a somewhat unfortunate drawback, in that they can make determining legal causation difficult. On the one hand,  the reduction methodology we define in \S \ref{HITL reductions} does give a consistent and formal mode of analysing HITL setups in order to assess the extent of the human's involvement in the computation, and the meaningfulness of their involvement. Particularly, where the human can influence computational systems through real queries, it can be said that they have more meaningful involvement. The consistency of this approach, and the level of formalism presented in \S \ref{HITL reductions} (and justified in \S \ref{practicalities} and \ref{ap_HITLtypes}), is also crucial for regulators in practice, enabling them to identify and address tokenistic HITL setups where the HITL has no meaningful involvement even though it may `appear' as though they are doing a lot.

On the other hand, in an \turing{}, it becomes extremely complex to determine whether the human actually influenced the machine's computation throughout the `ping-pong' process. Even if interaction logs are kept to trace the queries and back-and-forth interactions, it may not be possible to evaluate the impact of the human's input on the output of the ADMS due to the indeterminism of the computation tree. Each human input changes what the ADMS does next, and each ADMS query changes how the human views the process and thus how they might answer. It may become impossible to discern how and where failures arise and what these features are attributable to.

\subsection{Mesothelioma cases as a solution for \turing{}}\label{ap_Mesothelioma}

In this manuscript, we raise a problem concerning the assignment of responsibility in an \turing{} setup. Here, we explore how the law can confront this challenge by looking back to previous cases which have dealt with similar `responsibility gaps' in different contexts. The UK courts have previously departed from established causality principles to compensate workers who had developed deadly mesothelioma from exposure to asbestos fibres across multiple employers~\citep{Barker.2006}. In this line of cases, claimants had worked for several employers. At any point in their employment, they could have been exposed to asbestos fibres, causing them to develop mesothelioma. However, even a single asbestos fibre can trigger mesothelioma. As such, it was medically impossible to pinpoint which employer exposed the claimant to the asbestos that caused the mesothelioma. Departing from common causation principles in the mesothelioma cases, the court held that all the employers could be \textit{jointly liable} if it could be proven that they `materially increased the risk of harm', rather than finding a direct causal link between the employers and the mesothelioma.

In order to distribute the liability jointly, each employer's contribution to the claimant's overall risk of contracting the disease was considered. Courts sometimes divided the liability in proportion to the duration of exposure, even though they might not have actually caused harm directly. For example, if a claimant worked at one employer for 30\% of their `total exposure time', that employer was liable for 30\% of the damages. This approach was confirmed in cases like~\citet{Barker.2006} and later modified by the UK's Compensation Act 2006, which allowed claimants to recover full compensation from a single employer, who could then seek contribution from others.

The reasoning in the mesothelioma cases has received significant criticism from some legal academics since these cases were fundamentally decided on public policy grounds, as opposed to legal principle \citep{morgan2003lost}. The judges in the House of Lords reasoned that it would be unfair not to compensate workers just because causation could not be  attributed according to the existing legal tests. However, as shown in this manuscript, there are similar compelling public policy grounds to avoid the `scapegoating' of the human where responsibility gaps emerge. For example, although decided in accordance with US law, the Uber case described  in \S \ref{Uber example} shows how the human can face injustice if the liability is solely attributed to them. The Uber case involved numerous failure modes which spanned across the taxonomy presented in this manuscript and beyond the control of the human (see \S \ref{Uber example}). As such, when HITL setups fail, liability must be more distributed to account for responsibility gaps.

The mesothelioma cases are a good foundation for judges deciding cases involving the entangled human--machine configurations in \turing{}s because they reflect the impossibility of determining causation. In principle, the judges could hold any of the contributors that materially increase the risk of harm in the HITL setup liable. In an \turing{} setup, it may be impossible to determine what actually might have influenced the harm, but a myriad of factors could materially increase the risk of harm. Such factors could include any feature of the system that significantly enhance the likelihood of HITL failure modes in \S \ref{Failure taxonomy}, such as lack of training of the human, overworking the human causing fatigue, or even extreme bias in the training data used for an ADMS. Drawing upon the law's approach to liability in mesothelioma cases, in the context of this manuscript, identifying contributors which `materially increase the risk of harm' can be used by the  courts to hold multiple actors (including the technology companies) accountable so that the human is not treated as a `scapegoat' solution in `moral crumple zones'  \citep{Elish.2019}. Such legal solutions also incentivise companies to actively design systems to reduce the risk of failure because of the high liability consequences.

\end{document}